\documentclass[np2,twoside,twocolumn,showpacs]{revtex4}
\usepackage{amssymb}
\usepackage{amsmath}
\usepackage{bm,multirow}
\usepackage[dvips]{graphicx}
\usepackage{dcolumn}
\usepackage{longtable}
\usepackage{color}
\usepackage[version = 4]{mhchem}
\usepackage{comment}

\newcommand{\vect}[1]{\mbox{\boldmath $#1$}}       
\newcommand{\subm}[1]{_{\mathrm{#1}}}  
\newcommand{\CRA}{CeRh$_2$As$_2$~}

\begin{document}
\title{Investigation of the hyperfine coupling constant of locally noncentrosymmetric heavy-fermion superconductor \CRA}

\author{Shiki \surname{Ogata}}
\email{ogata.shiki.86c@st.kyoto-u.ac.jp}

\author{Shunsaku \surname{Kitagawa}, Mayu \surname{Kibune}, Kenji \surname{Ishida}}
\affiliation{Department of Physics, Kyoto University, Kyoto 606-8502, Japan}

\author{Katsuki \surname{Kinjo}}
\affiliation{Institute of Multidisciplinary Research for Advanced Materials, Tohoku University, Sendai, Miyagi 980-8577, Japan}

\author{Manuel \surname{Brando}, Christoph \surname{Geibel}, Seunghyun \surname{Khim}}
\affiliation{Max Planck Institute for Chemical Physics of Solids, D-01187 Dresden, Germany}

\author{Elena \surname{Hassinger}}
\affiliation{Max Planck Institute for Chemical Physics of Solids, D-01187 Dresden, Germany}
\affiliation{Technical University Dresden, Institute for Solid State and Materials Physics, 01062 Dresden, Germany}


\begin{abstract}
We performed $^{75}$As-NMR measurements in $H\parallel ab$ to investigate the normal-state magnetic properties of CeRh$_2$As$_2$, a recently-discovered heavy-fermion superconductor. We compared the NMR Knight shift $K$ with the magnetic susceptibility $\chi_{ab}$, and estimated the hyperfine coupling constant $A\subm{hf}$ from the slope of the $K-\chi$ plot. We observed that the magnitude of $A_{\mathrm{hf},ab}$ at the As(1) site changes at around 20 K owing to emerging the heavy-fermion state, which was also observed in $A\subm{hf}$ at the As(2) site and in $H\parallel c$. The sign of $A_{\mathrm{hf},ab}$ at the As(1) site is negative in low temperature. These are important for the analysis of the NMR results of CeRh$_2$As$_2$ in the superconducting state.
\end{abstract}


\maketitle

\section{INTRODUCTION}
In conventional superconductors, a superconducting order parameter can be classified as the multiplication of the states of spin $\vect{s}$ and momentum $\vect{k}$, that is, even-parity spin-singlet and odd-parity spin-triplet states. Recently, superconductors with the additional degrees of freedoms have been considered, such as frequency, atomic orbital state and sublattice \cite{odd-frequency,SRO_orbital,T.Yoshida}. In such systems, the novel superconducting states of odd-parity spin-singlet and even-parity spin-triplet states can be realized, because the fermionic antisymmetry should be held even in an additional degrees of freedom.\par 
\CRA is a recently discovered heavy-fermion superconductor, whose $T\subm{SC}$ is approximately 0.3 K \cite{hakken}. \CRA has the crystal structure of the tetragonal CaBe$_2$Ge$_2$-type with space group $P4/nmm$ (No.129, $D^{7}_{4h}$). \CRA has two crystallographically inequivalent As and Rh sites; As(1) [Rh(1)] is tetrahedrally coordinated by Rh(2) [As(2)], as shown in Fig.1(a). The heavy-fermion superconductivity is characterized by a broad maximum in resistivity at $T\subm{coh} \sim 40$ K and a large specific heat jump at $T\subm{SC}$. In the magnetic field parallel to the $c$-axis, the superconducting critical field $H\subm{c2}$ shows a clear kink at $H^* \sim 4$ T, and magnetic anomalies also appear at $H^*$ inside the superconducting phase, which suggests the superconducting multiphase. The high-field superconducting phase has a very large upper critical field $H\subm{c2} \sim 14$ T. For this superconducting multiphase, the local inversion-symmetry breaking is considered to be important. Theoretical study predicted that low-field even-parity and high-field odd-parity superconducting phases would be realized in locally inversion-breaking superconductors\cite{T.Yoshida}.\par
Along with this superconducting multiphase, it was reported that \CRA shows non-magnetic and magnetic orders just above and below $T\subm{SC}$, respectively \cite{hakken,QDW,QDW_theta,QDW_new,QDW_Los,Kibune,Ogata_c}. The specific heat shows a relatively small kink at $T_{0} \sim 0.4$ K together with large peak at $T\subm{SC}$, and the resistivity shows an upturn at $T_0$ \cite{QDW_new,QDW_Los} without the magnetic anomaly at $T_0$. $T_0$ is suppressed in $H\parallel c$, while it increases in $H\parallel ab$ with transition to another non-magnetic ordered states \cite{QDW}. Based on the field responses of $T_0$ \cite{QDW_theta,QDW_new}, $T_0$ is considered to be decoupled from the superconducting multiphase.\par
In addition to $T_0$, we have reported antiferromagnetic order inside the superconducting state from nuclear quadrupole resonance (NQR) \cite{Kibune} and nuclear magnetic resonance (NMR) \cite{Ogata_c} measurements. This antiferromagnetic order breaks the inversion symmetry. Remarkably, the antiferromagnetic order disappears almost simultaneously with $H^*$, which suggests the relation with the superconducting multiphase. Therefore, \CRA is a promising compound to study how the breaking of local inversion symmetry induces or influences unconventional nonmagnetic, antiferromagnetic and superconducting states, as well as their interaction.\par
It is important to investigate the spin state of the Cooper pair for understanding unconventional superconductors. The Knight-shift measurements using NMR provide critical information about the spin state. To estimate the spin susceptibility from the Knight shift measurement, it is necessary to estimate the hyperfine coupling constant $A\subm{hf}$. Previously, $A\subm{hf}$ has been estimated for two As sites in $H\parallel c$ and one As site [As(2) site] in $H\parallel ab$. However, $A\subm{hf}$ for the As(1) site in $H\parallel ab$ could not been estimated due to the limitations of the previous experimental setup \cite{Kibune_normal}.\par
In this study, we report the $^{75}$As-NMR results of the As(1) site in $H\parallel ab$. We compared the NMR Knight shift $K$ with the magnetic susceptibility $\chi_{ab}$, and estimated the hyperfine coupling constant $A\subm{hf}$ from the slope of the $K-\chi$ plot. It was evaluated that $A_{\mathrm{hf},ab}$ at the As(1) site is 0.24 T/$\mu\subm{B}$ above 20 K, and -0.22 T/$\mu\subm{B}$ below 20 K. The change in $A\subm{hf}$ is attributed to the development of the heavy-fermion state due to the $c-f$ hybridization. Negative $A\subm{hf}$ results in an increase in the Knight shift when the spin susceptibility decreases by formation of the spin-singlet pairs in superconducting state. This result is important to evaluate the spin susceptibility in the superconducting state of $H\parallel ab$.\par
   \begin{figure}[htbp!]
   \begin{center}
   \includegraphics[scale=0.89]{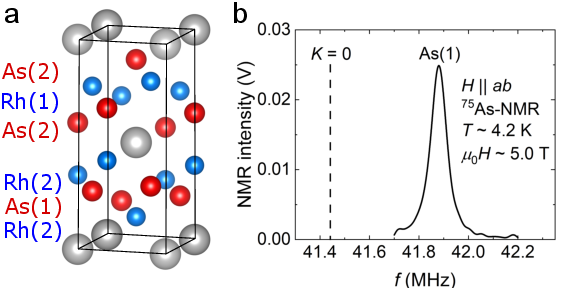}
   \end{center}
   \caption{(a)Crystal structure of CeRh$_2$As$_2$. (b)NMR spectrum of the As(1) site in $\mu_{0}H_{ab} \sim 5.0$ T. The dashed line indicate the position of $K=0$ when the contribution of the nuclear quadrupole interaction is taken into account.}
   \end{figure}
  \begin{table*}[hbtp!]
  \caption{Estimated hyperfine coupling constants $A\subm{hf}$ and $K\subm{orb}$. Temperatures in parentheses indicate the fitted temperature range. The data of Ref. \cite{Kibune_normal} is also listed.}
  \centering
   \begin{tabular}{ccccc}
   \hline
   Site & Direction of $H$ &
   \begin{tabular}{c}
   $A\subm{hf}$ $\left(\mathrm{T}/\mu\subm{B}\right)$\\ in low $T$
   \end{tabular}
   &  
   \begin{tabular}{c}
    $A\subm{hf}$ $\left(\mathrm{T}/\mu\subm{B}\right)$\\ in high $T$  
   \end{tabular}
   &
   \begin{tabular}{c}
    $K\subm{orb}$ $(\% )$\\ in high $T$  
   \end{tabular}
   \\
   \hline
   As(1)  & $H\parallel c$ & 
   \begin{tabular}{c}
   1.55 \\ (2-100 K)
   \end{tabular}
   & 
   \begin{tabular}{c}
   0.46 \\ (125-200 K)
   \end{tabular}
   &
   \begin{tabular}{c}
   0.57 \\ (125-200 K)
   \end{tabular}
   \\
     & 
     \begin{tabular}{c}
     $H\parallel ab$ \\ (New)
     \end{tabular}
     & 
   \begin{tabular}{c}
   -0.22 \\ (4.2-20 K)
   \end{tabular}
   & 
   \begin{tabular}{c}
   0.24 \\ (20-50 K)
   \end{tabular}
   &
   \begin{tabular}{c}
   0.91 \\ (20-50 K)
   \end{tabular}
   \\
   \hline
   As(2)  & $H\parallel c$ &
   \begin{tabular}{c}
   0.27\\(2-30 K)
   \end{tabular}
   & 
   \begin{tabular}{c}
   0.55\\(40-200 K)
   \end{tabular}
   &
   \begin{tabular}{c}
   0.10\\(40-200 K)
   \end{tabular}\\
    & $H\parallel ab$ & 
   \begin{tabular}{c}
   0.16\\(2-10 K)
   \end{tabular}
   & 
   \begin{tabular}{c}
   0.50\\(15-200 K)
   \end{tabular}
   &
   \begin{tabular}{c}
   0.09\\(15-200 K)
   \end{tabular}\\
   \hline
  \end{tabular}
 \end{table*}
 
\section{EXPERIMENTS}
Single crystals of \CRA were grown using the bismuth flux method \cite{hakken}. The bulk magnetic susceptibility was measured using a commercial magnetometer with a superconducting quantum interference device (Quantum Design, MPMS). A conventional spin-echo technique was used for the NMR measurements. The $^{75}$As-NMR spectra (nuclear spin $I$ = 3/2, nuclear gyromagnetic ratio $\gamma$/2$\pi$ = 7.29 MHz/T, and natural abundance 100\% ) were obtained as a function of frequency at fixed magnetic fields ($\sim$ 5.0 T). We observed the As(1) site around $f\sim 41.9$ MHz at $\mu_{0}H\sim 5.0$ T, as shown in Fig. 1(b). The site assignment of the NMR peaks has been described in a previous paper \cite{Kibune_normal}. To estimate the Knight shift $K_{i}$ ($i = c$ and $ab$), we computed the resonance frequency by diagonalizing the following nuclear Hamiltonian; 
\begin{flalign}
    \mathcal{H} &= -\left(\frac{\gamma}{2\pi}\right) h (1+K_{i}) \vect{I}\cdot \vect{H}\nonumber\\
    &+ \frac{h\nu_{Q}}{6}\left[3I^{2}_{z} - I(I+1) + \frac{\eta}{2}(I^{2}_{+} + I^{2}_{-})\right],
\end{flalign}
where $h$, $\nu_{Q} = \frac{3eQV_{zz}}{2I(2I-1)}$, and $\eta = |\frac{V_{yy}-V_{xx}}{V_{zz}}|$ are the Planck constant, NQR frequency, and asymmetry parameter, respectively. $\eta$ is zero at each As site because of the 4-fold symmetry of the atomic position. The temperature dependence of $\nu_{Q}$ was determined from NQR measurements. The magnetic field was calibrated using $^{63}$Cu ($^{63}\gamma _n$/2$\pi$ = 11.289 MHz/T) and $^{65}$Cu ($^{65}\gamma _n$/2$\pi$ = 12.093 MHz/T) NMR signals from the NMR coil.\par

\section{RESULTS AND DISCUSSION}
Figure 2 shows the temperature dependence of the Knight shift $K_{ab}$ at the As(1) site (measured in this study) and the As(2) site (measured in previous study \cite{Kibune_normal}). For comparison, the temperature dependence of the magnetic susceptibility $\chi_{ab}$ is also shown. The Knight shift is estimated from the peak frequency of the Gaussian fittings. The Knight shift is proportional to the magnetic susceptibility as given by,
\begin{flalign}
    K_{i} = A_{\mathrm{hf},i} \chi_{\mathrm{spin},i} + K_{\mathrm{orb},i},
\end{flalign}
where $A_{\mathrm{hf},i}$, $\chi_{\mathrm{spin},i}$, and $K_{\mathrm{orb},i}$ are the hyperfine coupling constant, spin susceptibility, and orbital part of the Knight shift in each direction, respectively. $K_{\mathrm{orb},i}$ is generally temperature independent. Above 20 K, $K_{ab}$ of both As sites increases on cooling, while $K_{ab}$ of the As(1) site shows broad maximum at 20 K. Such behaviors are attributed to the development of the heavy-fermion state due to the $c-f$ hybridization, which are often observed in Ce-based heavy-fermion systems \cite{c-f_hybridization1,c-f_hybridization2}. Indeed, the temperature where the $K_{ab}$ shows the broad maximum is roughly match $T\subm{coh} \sim 40$ K estimated from the resistivity measurement.\par
   \begin{figure}[htbp!]
   \begin{center}
   \includegraphics[scale=1.5]{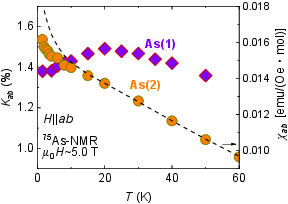}
   \end{center}
   \caption{The temperature dependence of the Knight shift in $ab$-plane field $K_{ab}$. For comparison, the temperature dependence of magnetic susceptibility $\chi_{ab}$ is also depicted (the dashed line).}
   \end{figure}
%
   \begin{figure}[htbp!]
   \begin{center}
   \includegraphics[scale=1.5]{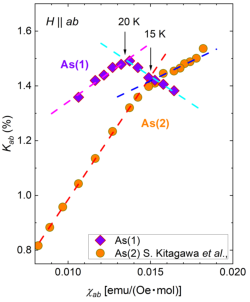}
   \end{center}
   \caption{$K - \chi$ plot for $H\parallel ab$. The dashed lines indicate the results of linear fitting.}
   \end{figure}
The hyperfine coupling constant $A_{\mathrm{hf},i}$ can be determined from the slope of the $K-\chi$ plot, as shown in Fig. 3. For both sites, the slopes of $K-\chi$ plot are changed between the regions with small and large $\chi\subm{AC}$. This behavior was observed also in $H\parallel c$. Curro \textit{et al.} suggested that the magnitude of $A\subm{hf}$ is changed at $T^*$, below which heavy-fermion state starts to developed \cite{curro2004scaling}. We obtained $A_{\mathrm{hf},ab}$ from fitting the slopes in each region. Table I summarizes the estimated hyperfine coupling constants $A\subm{hf}$ and $K\subm{orb}$ by this study and \cite{Kibune_normal}. $A_{\mathrm{hf},c}$ at the As(1) site becomes larger on cooling, while $A_{\mathrm{hf},c}$ at the As(2) site and $A_{\mathrm{hf},ab}$ at both As sites become smaller on cooling. Notably, $A_{\mathrm{hf},ab}$ at the As(1) site becomes negative in low temperature region. Assuming the classical dipole interaction, $A_{\mathrm{hf},i}$ was calculated to be 0.02-0.06 T/$\mu\subm{B}$ \cite{Kibune_normal}, which is of one order of magnitude smaller than the experimentally estimated $A_{\mathrm{hf},i}$. Therefore, the anomalous transferred hyperfine field is dominant in CeRh$_2$As$_2$. It is noteworthy that the similar change in the sign of $A\subm{hf}$ was observed in $A_{\mathrm{hf},c}$ at the Si site in CeCu$_2$Si$_2$ \cite{c-f_hybridization1} and in $A_{\mathrm{hf},b}$ at the In(2) site in CeCoIn$_5$ \cite{c-f_hybridization2}. Although the obvious change in $A\subm{hf}$ at $T^*$ was observed in most of Ce-based heavy-fermion compounds, it seems that the mechanism and the estimation of the change have not been well understood.\par
\section{CONCLUSION}
In conclusion, we performed $^{75}$As-NMR measurements to investigate the hyperfine coupling constant $A_{\mathrm{hf},ab}$ at the As(1) site of CeRh$_2$As$_2$, which was not measured in previous study \cite{Kibune_normal}. $A_{\mathrm{hf},ab}$ at the As(1) site changed from positive (0.24 T/$\mu\subm{B}$) to negative (-0.22 T/$\mu\subm{B}$) at around 20 K due to the formation of the heavy-fermion state; changes in $A\subm{hf}$ have been observed at other sites and in $H\parallel c$. These results are essential for NMR studies of \CRA in $H\parallel ab$, which is necessary to determine the magnetic structure of antiferromagnetic state, and the decrease of the spin susceptibility in the superconducting state.

\section*{ACKNOWLEDGEMENTS}
This work was partially supported by the Kyoto University LTM Center and Grants-in-Aid for Scientific Research (KAKENHI) (Grants No. JP19K14657, No. JP19H04696, No. JP20KK0061, No. JP20H00130, No. JP21K18600, No. JP22H04933, No. JP22H01168, and No.JP23H01124).  C. G. and E. H. acknowledge support from the DFG program Fermi-NESt through Grant No. GE 602/4-1. Additionally, E. H. acknowledges funding by the DFG through CRC1143 (Project No. 247310070) and the Würzburg-Dresden Cluster of Excellence on Complexity and Topology in Quantum Matter—ct.qmat (EXC 2147, Project ID 390858490). S. O. would like to acknowledge the support from the Motizuki Fund of Yukawa Memorial Foundation.

%

\end{document}